\documentstyle[aps,12pt]{revtex}
\begin{document}
\author{J.R. Morris}
\address{{\it Physics Department, Indiana University Northwest, 3400 Broadway, Gary,}%
\\
Indiana 46408}
\author{D. Bazeia}
\address{{\it Departamento de F\'{\i}sica, Universidade Federal da Para\'{\i}ba,}\\
{\it Caixa Postal 5008, 58051-970 Jo\~{a}o Pessoa, Para\'{\i}ba, Brazil}\\
\bigskip\ }
\title{Supersymmetry Breaking and Fermi Balls}
\maketitle

\begin{abstract}
A simple model is presented where the disappearance of domain walls and the
associated production of ``Fermi balls'', which have been proposed as
candidates for cold dark matter, are features which arise rather naturally
in response to softly broken supersymmetry.

\smallskip\ 

PACS:  11.27.+d, 12.60.Jv, 98.80.Cq
\end{abstract}

\section{Introduction}

The possibility exists that the early universe experienced a sequence of
symmetry breaking phase transitions, which could have resulted in the
production of defects, such as monopoles, cosmic strings, or domain walls 
\cite{vilenkin,vsbook,ktbook}. Furthermore, it is possible that
supersymmetry could have been physically realized during early epochs,
becoming broken after defect formation, at a lower energy scale. It is
therefore quite natural to investigate models of defects within a
supersymmetric context. Here, attention is focused upon a simple
supersymmetric model constructed from a single chiral supermultiplet, which
admits a domain wall solution interpolating between two distinct, but
energetically degenerate, supersymmetric vacuum states. The domain wall
formation arises from an exact discrete symmetry which is spontaneously
broken. The initial supersymmetry of the model couples the fermion fields to
the scalar fields in a prescribed way, and it is found that, as a result of
this coupling, a fermion zero mode\cite{jr} forms within the core of the
domain wall, where the fermion is essentially massless. Bound states can
also exist which describe scalar bosons attached to the wall.

The breaking of supersymmetry at lower energies can be described by the
inclusion of soft supersymmetry breaking terms in the scalar potential.
However, it is found that when the soft supersymmetry breaking terms are
added to the Lagrangian, the {\it exact} discrete symmetry responsible for
the formation of the stable domain wall is transformed into an {\it %
approximate}, or {\it biased}, discrete symmetry. This approximate discrete
symmetry results in a domain wall network where each wall interpolates
between two different, energetically nondegenerate (and nonsupersymmetric)
vacuum states -- a true vacuum state and a higher energy density false
vacuum state\cite{ggk,fggk}. Thus, the terms that are added to the
Lagrangian to break the supersymmetry also explicitly break the exact
discrete symmetry. The model then resembles one recently proposed by
Macpherson and Campbell (MC)\cite{mc} wherein a biased discrete symmetry
breaking results in the production of ``Fermi balls'' -- tiny bags of false
vacuum that are inhabited by a stabilizing Fermi gas. The Fermi balls emerge
as an end product of the collapse and fragmentation of domain walls
enclosing false vacuum protodomains. Massive, electrically neutral Fermi
balls can be considered as candidates for cold dark matter. The simple model
presented here therefore connects, in a rather natural way, the
transformation of domain walls into Fermi balls and the breaking of
supersymmetry.

The supersymmetric model is presented in the next section, where a domain
wall solution is found. Upon examining the response of the fermion and boson
fields to the domain wall background, it is seen that a fermion zero mode
forms inside the wall, and that there can exist bound states describing the
attachment of bosons to the wall. The fermion mass vanishes in the core of
the wall, and fermions near the wall can experience a short ranged, but
strong, attractive force toward the core of the wall where it is
energetically more favorable for them to reside. This effect will be of
importance in the consideration of Fermi balls, since the wall will tend to
absorb fermions to become populated by a Fermi gas of effectively massless
fermions that can contribute a fermion degeneracy pressure. The soft
supersymmetry breaking terms that are to be added to the Lagrangian are
given in sec. III. It is easily seen that the inclusion of these terms
causes the previous exact discrete symmetry associated with the formation of
the domain wall to be exchanged for an approximate, biased discrete
symmetry. The basic mechanisms proposed by MC relating to the disappearance
of the domain walls through the formation of false vacuum bags, the collapse
and fragmentation of the vacuum bags, and the production of Fermi balls are
briefly reviewed. It is seen that the model presented here, with the
breaking of supersymmetry, closely resembles the model presented for Fermi
balls, allowing an inference that, at least in the context of the simple
model presented here, the production of Fermi balls can arise from the
breaking of supersymmetry and a discrete symmetry. A short summary forms
sec. IV.

\section{The Supersymmetric Model}

Consider a supersymmetric model constructed from a single chiral superfield $%
\Phi $ with component fields $(\phi ,\psi ,F)$, where $F$ represents the
auxiliary boson field. The boson fields $\phi $ and $F$ are complex scalar
fields and the fermion field $\psi $ is a Weyl two-spinor. Let us write the
scalar field $\phi $ in the form $\phi =A+iB$, where $A$ and $B$ are
real-valued. The superfield $\Phi $ has a superspace representation\cite
{sbook,wbbook} given by 
\begin{equation}
\Phi (y,\theta )=\phi (y)+\sqrt{2}\theta \psi (y)+\theta ^2F(y),  \label{e1}
\end{equation}

\noindent where $y^\mu =x^\mu +i\theta \sigma ^\mu \bar{\theta}$ and $\theta
^2=\theta \theta =\theta ^\alpha \theta _\alpha $, $\alpha =1,2$. (We also
have $\theta \psi =\theta ^\alpha \psi _\alpha $, $\bar{\theta}\bar{\psi}=%
\bar{\theta}_{\dot{\alpha}}\bar{\psi}^{\dot{\alpha}}$, $\dot{\alpha}=1,2$.)
A metric $g_{\mu \nu }$ with signature $(+,-,-,-)$ is used. (See the
Appendix for a brief description of the conventions and gamma matrices.) One
can also define a Majorana 4-spinor $\Psi $ in terms of the Weyl 2-spinors: 
\begin{equation}
\Psi =\left( 
\begin{array}{l}
\psi _\alpha \\ 
\bar{\psi}^{\dot{\alpha}}
\end{array}
\right) ,\,\,\,\,\,\,\,\,\,\,\alpha =1,2,\,\,\,\,\,\dot{\alpha}=1,2.
\label{e2}
\end{equation}

\subsection{Lagrangian}

In terms of the chiral superfield the Lagrangian can be written as 
\begin{equation}
L=(\bar{\Phi}\Phi )|_{\theta ^2\bar{\theta}^2}+W(\Phi )|_{\theta ^2}+\bar{W}(%
\bar{\Phi})|_{\bar{\theta}^2}\,\,,  \label{e3}
\end{equation}

\noindent where $\bar{\Phi}=\Phi ^{*}$, i.e., the ``bar'' and ``star''
symbols mean complex conjugation, $W(\Phi )$ is the superpotential, which
will be defined shortly, and $X|_{\theta ^2}$ stands for the $\theta ^2$
part of $X$, etc. By eliminating the auxiliary field $F$, the Lagrangian can
be written in terms of the component fields as 
\begin{equation}
L=L_K^B+L_K^F+L_Y-V,  \label{e4}
\end{equation}

\noindent where 
\begin{equation}
L_K^B=\partial ^\mu \bar{\phi}\partial _\mu \phi =\partial ^\mu A\partial
_\mu A+\partial ^\mu B\partial _\mu B,\,\,\,\,\,\,\,\,\,\phi =A+iB,\,
\label{e5}
\end{equation}
\begin{equation}
L_K^F=%
{\textstyle {i \over 2}}
\left[ (\partial _\mu \psi )\sigma ^\mu \bar{\psi}-\psi \sigma ^\mu \partial
_\mu \bar{\psi}\right] =%
{\textstyle {i \over 2}}
\bar{\Psi}\gamma ^\mu \partial _\mu \Psi ,  \label{e6}
\end{equation}
\begin{equation}
L_Y=-%
{\textstyle {1 \over 2}}
\left[ \left( \frac{\partial ^2W}{\partial \phi ^2}\right) \psi \psi +\left( 
\frac{\partial ^2W}{\partial \phi ^2}\right) ^{*}\bar{\psi}\bar{\psi}\right]
,  \label{e7}
\end{equation}
\begin{equation}
V=|F|^2=\left| \frac{\partial W}{\partial \phi }\right|
^2,\,\,\,\,\,\,\,\,\,\,F=-\left( \frac{\partial W}{\partial \phi }\right)
^{*}\,\,.  \label{e8}
\end{equation}

\subsection{Superpotential and Scalar Potential}

To get a domain wall solution, let us choose the superpotential 
\begin{equation}
W=\lambda \Phi \left( 
{\textstyle {1 \over 3}}
\Phi ^2-a^2\right)  \label{e9}
\end{equation}

\noindent so that the auxiliary field is described by $F^{*}=-\lambda (\phi
^2-a^2)$ , where $a$ is a constant. From (\ref{e8}) the scalar potential is
then given by 
\begin{equation}
V=F^{*}F=\lambda ^2(\bar{\phi}^2-a^2)(\phi ^2-a^2)=\lambda ^2[(\bar{\phi}%
\phi )^2-a^2(\phi ^2+\bar{\phi}^2)+a^4].  \label{e10}
\end{equation}

\noindent The scalar potential $V=|F|^2\geq 0$ has minima located at $F=0$,
which implies that the (supersymmetric) vacuum states of the theory are
located by $\phi =\pm a$, i.e., the vacuum states of the theory are
described by $A=\pm a$, $B=0$. It is seen that supersymmetry is respected in
the vacuum, where $V=0$.

\subsection{Field Equations}

Recalling that $\phi =A+iB$, along with the expression for the Majorana
spinor $\Psi $ given by (\ref{e2}), the Lagrangian, expressed in terms of
the real scalar fields $A$ and $B$ and the Majorana 4-spinor $\Psi $, can be
written in the form given by (\ref{e4}) with (using $\partial ^2W/\partial
\phi ^2=2\lambda \phi $) 
\begin{equation}
\begin{array}{ll}
L & =\partial ^\mu A\partial _\mu A+\partial ^\mu B\partial _\mu B+\frac i2%
\bar{\Psi}\gamma ^\mu \partial _\mu \Psi +i\lambda \left[ A\bar{\Psi}\Psi +B%
\bar{\Psi}\gamma _5\Psi \right] \\ 
& -\lambda ^2\left[ (A^2-a^2)^2+2B^2(A^2+a^2)+B^4\right] ,
\end{array}
\label{e11}
\end{equation}

\noindent where ($\psi \psi +\bar{\psi}\bar{\psi})=-i\bar{\Psi}\Psi $ and $%
(\psi \psi -\bar{\psi}\bar{\psi})=-\bar{\Psi}\gamma _5\Psi $ have been used.
With $L=L_K^B+L_K^F+L_Y-V$, the field equations for $A$, $B$, and $\Psi $
follow from 
\begin{equation}
2\Box A+\frac{\partial V}{\partial A}-\frac{\partial L_Y}{\partial A}=0,
\label{e12}
\end{equation}
\begin{equation}
2\Box B+\frac{\partial V}{\partial B}-\frac{\partial L_Y}{\partial B}=0,
\label{e13}
\end{equation}
\begin{equation}
\frac{\partial L}{\partial \bar{\Psi}}=\frac{\partial L_K^F}{\partial \bar{%
\Psi}}+\frac{\partial L_Y}{\partial \bar{\Psi}}=0.  \label{e14}
\end{equation}

\noindent Therefore, from (\ref{e11})-(\ref{e14}), the field equations for $%
A $, $B$, and $\Psi $ are given by 
\begin{equation}
\Box A+2\lambda ^2A(A^2+B^2-a^2)-i%
{\textstyle {\lambda \over 2}}
\bar{\Psi}\Psi =0,  \label{e15}
\end{equation}
\begin{equation}
\Box B+2\lambda ^2B(A^2+B^2+a^2)-i%
{\textstyle {\lambda \over 2}}
\bar{\Psi}\gamma _5\Psi =0,  \label{e16}
\end{equation}
\begin{equation}
\gamma ^\mu \partial _\mu \Psi +2\lambda (A+B\gamma _5)\Psi =0,  \label{e17}
\end{equation}

\noindent where $\Box =\partial ^\mu \partial _\mu $.

\subsection{The Domain Wall and Particle Masses}

Let us consider the real bosonic sector of the model where the fermionic
fields vanish and the scalar field is real-valued, i.e. $\Psi =0$, $B=0$. As
boundary conditions for the $A$ field we take $A(x=\pm \infty )=\pm a$. Then
(\ref{e15}) gives 
\begin{equation}
\Box A+2\lambda ^2A(A^2-a^2)=0,  \label{e18}
\end{equation}

\noindent which has as a static solution 
\begin{equation}
A_W(x)=a\tanh \frac xw,\,\,\,\,\,\,\,\,\,\,w=\frac 1{\lambda a}.  \label{e19}
\end{equation}

The solution given by (\ref{e19}) describes an ordinary domain wall of
thickness $w=\frac 1{\lambda a}$. We can notice that the domain wall can be
associated with the spontaneous breaking of an exact discrete $Z_2$ symmetry
describing invariance of the real bosonic Lagrangian under $A\rightarrow -A$%
. Using this domain wall solution as a background solution, the response of
the fields $\Psi $ and $B$ can be examined. It will be seen that there is a
fermionic zero mode within the domain wall, and that there are domain wall-$%
B $ particle bound states.

In the vacuum states we have $A=\pm a$, $B=0$, $\Psi =0$. Therefore, in
vacuum the $B$ particle mass is determined to be $m_B=2\lambda a$, and for
the Majorana fermion, $L_Y^{(+a)}=i\lambda a\bar{\Psi}\Psi =\frac i2m_F\bar{%
\Psi}\Psi $, which gives $m_F=2\lambda a=m_B$, which is expected from
supersymmetry in the vacuum. (For the vacuum state $A=-a$, the mass
eigenstate Weyl spinors must be redefined by a phase rotation, and the
Majorana spinor undergoes a $\gamma _5$ rotation, $\Psi \rightarrow \gamma
_5\Psi $.)

In the core of the domain wall, $A\rightarrow 0$, and we find $m_F=0$, $m_B=%
\sqrt{2}\lambda a$, so that the mass of each particle decreases within the
domain wall. On this basis, we see that the particles are attracted toward
the wall with a force $F\sim -\partial m(x)/\partial x$. The existence of
scalar bound states and spinor zero modes is consistent with this picture.
For the Majorana fermion we have $m_F(x)=2\lambda A(x)$ (for $x>0$, e.g.)
and therefore, by (\ref{e19}), the force of attraction can be estimated to
be $F\sim -\frac 2{w^2}\left[ \cosh \frac xw\right] ^{-2}$, which, for a
thin wall, can be quite large in magnitude (but of short range, rapidly
vanishing outside the wall's surface).

\subsection{Fermionic Zero Mode}

\subsubsection{Static Zero Mode}

Upon setting $A=A_W(x)$, $B=0$, the field equation for $\Psi $ reduces to 
\begin{equation}
\gamma ^\mu \partial _\mu \Psi +2\lambda A_W\Psi =0.  \label{e20}
\end{equation}

\noindent For the gamma matrices we have $\left\{ \gamma ^\mu ,\gamma ^\nu
\right\} =-2g^{\mu \nu }$, $\,\gamma ^1=i\left( 
\begin{array}{ll}
0 & \sigma _1 \\ 
-\sigma _1 & 0
\end{array}
\right) $, $(\gamma ^1)^2=1$. Let us first look for a static solution of the
form $\Psi =\Psi (x)$. Multiplying (\ref{e20}) by $\gamma ^1$ gives 
\begin{equation}
\partial _x\Psi =-2\lambda A_W\gamma ^1\Psi .  \label{e21}
\end{equation}

\noindent Let us now write the Majorana 4-spinor $\Psi $ in terms of
2-spinors $\eta $ and $\chi $: $\Psi =\left( 
\begin{array}{l}
\eta \\ 
\chi
\end{array}
\right) $. We then have $\gamma ^1\Psi =i\left( 
\begin{array}{ll}
0 & \sigma _1 \\ 
-\sigma _1 & 0
\end{array}
\right) \left( 
\begin{array}{l}
\eta \\ 
\chi
\end{array}
\right) =i\left( 
\begin{array}{l}
\sigma _1\chi \\ 
-\sigma _1\eta
\end{array}
\right) $. Therefore, 
\begin{equation}
\partial _x\left( 
\begin{array}{l}
\eta \\ 
\chi
\end{array}
\right) =-2i\lambda A_W\left( 
\begin{array}{l}
\sigma _1\chi \\ 
-\sigma _1\eta
\end{array}
\right) ,\,\,\,\,\,\,\,\,\,\,\sigma _1=\left( 
\begin{array}{ll}
0 & 1 \\ 
1 & 0
\end{array}
\right) ,\,\,\,\,\,\,\,(\sigma _1)^2=1.  \label{e22}
\end{equation}

The equations for $\eta $ and $\chi $ can be decoupled by writing 
\begin{equation}
\chi =-i\sigma _1\eta ,\,\,\,\,\,\,\,\,\,\,\eta =i\sigma _1\chi .
\label{e23}
\end{equation}

\noindent Then, by (\ref{e22}) and (\ref{e23}), 
\begin{equation}
\partial _x\eta =-2\lambda A_W\eta ,\,\,\,\,\,\,\,\,\,\,\partial _x\chi
=-2\lambda A_W\chi ,\,\,\,\,\,\,\,\,\,\,\Psi =\left( 
\begin{array}{l}
\eta \\ 
\chi
\end{array}
\right) =\left( 
\begin{array}{l}
\eta \\ 
-i\sigma _1\eta
\end{array}
\right) .  \label{e24}
\end{equation}

\noindent A solution is given by 
\begin{equation}
\eta =\tau \exp \left[ -2\lambda \int_0^xA_W(x^{\prime })dx^{\prime }\right]
=\tau \left[ \cosh \frac xw\right] ^{-2},  \label{e25}
\end{equation}

\noindent where $\tau $ is an arbitrary constant Weyl 2-spinor.

The Majorana condition $\Psi _C=-\gamma ^2\Psi ^{*}=\Psi $, (where $\Psi _C$
is the charge conjugate of $\Psi )$ i.e. 
\begin{equation}
\Psi =\left( 
\begin{array}{l}
\eta \\ 
\chi
\end{array}
\right) =\left( 
\begin{array}{l}
\eta \\ 
i\sigma _2\eta ^{*}
\end{array}
\right) ,  \label{e26}
\end{equation}

\noindent must also be satisfied. Upon comparing (\ref{e24}) and (\ref{e26}%
), we have $\sigma _2\eta ^{*}=-\sigma _1\eta $, or $\sigma _1\sigma _2\eta
^{*}=-\eta $, so that with the help of $\sigma _1\sigma _2=i\sigma _3$, we
get $\eta ^{*}=i\sigma _3\eta $. We must therefore require that $\tau
^{*}=i\sigma _3\tau $.

\subsubsection{Traveling Waves}

Let us now regard $\Psi $ to be a function of $x$, $z$, and $t$, i.e., $\Psi
(x,z,t)=\left( 
\begin{array}{l}
\eta (x,z,t) \\ 
-i\sigma _1\eta (x,z,t)
\end{array}
\right) $, where $\eta (x,z,t)=\tau (z,t)\left[ \cosh \frac xw\right] ^{-2}$%
. Then (\ref{e20}) implies that 
\begin{equation}
(\gamma ^0\partial _0+\gamma _3\partial _3)\left( 
\begin{array}{l}
\tau (z,t) \\ 
-i\sigma _1\tau (z,t)
\end{array}
\right) \left[ \cosh \frac xw\right] ^{-2}=0,  \label{e27}
\end{equation}

\noindent which is solved by 
\begin{equation}
(\partial _0-\sigma _3\partial _3)\tau (z,t)=0.  \label{e28}
\end{equation}

\noindent This can be seen by multiplying (\ref{e27}) by $\gamma ^1$ and
using $\gamma ^0\gamma ^3=\left( 
\begin{array}{ll}
\sigma _3 & 0 \\ 
0 & -\sigma _3
\end{array}
\right) $, so that (\ref{e27}) reduces to the set of equations $(\partial
_0-\sigma _3\partial _3)\tau =0,$ and $(\partial _0+\sigma _3\partial
_3)\sigma _1\tau =0$, and the second equation is automatically solved when
the first equation is solved. Then (\ref{e28}) can be written explicitly as 
\begin{equation}
\left( 
\begin{array}{ll}
(\partial _0-\partial _3) & 0 \\ 
0 & (\partial _0+\partial _3)
\end{array}
\right) \left( 
\begin{array}{l}
\tau _1(z,t) \\ 
\tau _2(z,t)
\end{array}
\right) =0.  \label{e29}
\end{equation}

\noindent This is solved by 
\begin{equation}
\tau _1(z,t)=\tau _1(z+t),\,\,\,\,\,\,\,\,\,\,\tau _2(z,t)=\tau _2(z-t).
\label{e30}
\end{equation}

\noindent Therefore, $\tau $ can be written as 
\begin{equation}
\tau (z,t)=\left( 
\begin{array}{l}
\tau _1(z+t) \\ 
0
\end{array}
\right) +\left( 
\begin{array}{l}
0 \\ 
\tau _2(z-t)
\end{array}
\right) ,  \label{e31}
\end{equation}

\noindent so that $\tau $, and hence $\Psi $, can contain a linear
combination of ``up'' and ``down'' moving waves.

\subsection{B Particle Bound States}

Now let us set $\Psi $ equal to zero and examine the $B$ field in the domain
wall background. From the field equation for $B$, we have 
\begin{equation}
\Box B+2\lambda ^2B(A_W^2+B^2+a^2)=0.  \label{e32}
\end{equation}

\noindent Now linearize, and look at small fluctuations about $B=0$ to
obtain 
\begin{equation}
\Box B+2\lambda ^2a^2\left( 1+\tanh ^2\frac xw\right) B=0.  \label{e33}
\end{equation}

\noindent Writing $B(x,z,t)=b(x)\sin (kz-\omega t+\delta )$, (\ref{e33})
reduces to 
\begin{equation}
-\partial _x^2b+2\lambda ^2a^2[\tanh ^2(x/w)]b=E^2b,\,\,\,\,\,\,\,E^2\equiv
\omega ^2-(k^2+2\lambda ^2a^2).  \label{e34}
\end{equation}

\noindent This is a Schrodinger-like equation with an attractive potential
that can accommodate one or more bound states\cite{mf} with $0<E<\sqrt{2}%
\lambda a$. We therefore infer that real scalar $B$ particles can be
localized within or near the core of the domain wall. For $E>\sqrt{2}\lambda
a$ there can exist a set of states describing the scattering of $B$
particles from the domain wall.

\section{Soft Supersymmetry Breaking and Fermi Balls}

The supersymmetry that exists in the vacuum states of the above model can be
broken by adding soft supersymmetry breaking terms to the scalar potential%
\cite{mohap}. The types of soft terms allowed here include a scalar mass
term of the form $\mu ^2\bar{\phi}\phi $ and a trilinear scalar interaction
of the form [$W(\Phi )|_{\theta =0}+c.c.]=g_0[\phi ^3+\bar{\phi}^3]$. Let us
also add a (dynamically irrelevant) constant $V_0$ and therefore define the
potential term 
\begin{equation}
\begin{array}{ll}
V_B & =\mu ^2\bar{\phi}\phi +g_0\left( \phi ^3+\bar{\phi}^3\right) +V_0 \\ 
& =\mu ^2(A^2+B^2)+g_0(A^3-3AB^2)+V_0,
\end{array}
\label{e35}
\end{equation}

\noindent where the constant $V_0$ can be used to adjust the vacuum energy
of the true vacuum state to zero. The total potential can be written as 
\begin{equation}
V_1=V+V_B.  \label{e36}
\end{equation}

\noindent Notice that not only has the original supersymmetry been broken,
but the discrete $Z_2$ symmetry associated with the reflection $A\rightarrow
-A$ has also been broken by $V_B$. In fact, the model now resembles the kind
of model\cite{mc} that was introduced for the description of a {\it Fermi
ball.} Therefore, we have the possibility that a simple supersymmetric
model, which possesses an exact discrete symmetry before the breaking of
supersymmetry, can be left with an approximate discrete symmetry after the
breaking of the supersymmetry, so that energetically nondegenerate vacuum
states develop -- a true vacuum state and a higher energy false vacuum
state. Two different domains are separated by a domain wall, which can
rapidly fragment into a mist of Fermi balls by the mechanism described by
MacPherson and Campbell (MC). Massive neutral Fermi balls which couple only
weakly with ordinary matter are then candidates for cold dark matter. Let us
briefly review the Fermi ball model proposed by MC and then examine the
model presented here to see how it can describe Fermi balls.

\subsection{Fermi Balls}

The basic scenario described by MC for the production of Fermi balls can be
briefly summarized in the following. (For discussions of biased discrete
symmetry breaking and domain walls, see also refs.\cite{ggk,fggk}.) Consider
a simple model of a self-interacting scalar field $\varphi $ and a Dirac
fermion field $\psi $ which is strongly coupled to the scalar field $\varphi 
$. The system can be described as follows: first consider the Lagrangian 
\begin{equation}
L_0=\frac 12\partial ^\mu \varphi \partial _\mu \varphi -\frac{\lambda ^2}8%
(\varphi ^2-\varphi _0^2)^2.  \label{e37}
\end{equation}

\noindent $L_0$ possesses a discrete $Z_2$ reflection symmetry, and the
degenerate vacuum states of the theory are described by $\varphi =\pm
\varphi _0$. A domain wall solution of the form $\varphi =\varphi _0\tanh
\left( \frac x\delta \right) $ interpolates between the two distinct vacuum
states, with $\varphi \rightarrow \pm \varphi _0$ as $x\rightarrow \pm
\infty $, and in the core of the domain wall $\varphi \rightarrow 0$. The
surface tension is equal to the surface energy density, given by 
\begin{equation}
\sigma =\frac{2\lambda \varphi _0^3}3.  \label{e38}
\end{equation}

Let us now consider changing the exact discrete $Z_2$ symmetry to an
approximate symmetry. The symmetry breaking results in the formation of two
different, nondegenerate, vacuum states that form protodomains of true and
false vacuum. Let the difference in the energy densities of the two vacuum
states be represented by $\Lambda $. In this biased discrete symmetry
breaking a domain wall can form which interpolates between the true and
false vacuum protodomains. The exact $Z_2$ symmetry can be exchanged for an
approximate symmetry by adding, for example, a $Z_2$ symmetry breaking term $%
A(\varphi )$ to the Lagrangian. The asymmetry that is introduced can result
in the formation of finite sized ``false vacuum bags'' -- regions of false
vacuum protodomain enclosed by domain wall\cite{ggk,fggk}. These vacuum bags
can collapse, and result in the conversion of false vacuum into true vacuum
and the disappearance of the domain walls.

Let us now consider a Dirac fermion $\psi $ that is strongly coupled to the
scalar field $\varphi $ through a standard Yukawa coupling. We replace the
Lagrangian $L_0$ with 
\begin{equation}
L=\bar{\psi}(i\gamma ^\mu \partial _\mu +iG\varphi )\psi +\frac 12\partial
^\mu \varphi \partial _\mu \varphi -\frac{\lambda ^2}8(\varphi ^2-\varphi
_0^2)^2+A(\varphi ).  \label{e39}
\end{equation}

\noindent [A relative factor of $(-i)$ appears in the Yukawa term in (\ref
{e39}) due to our choice of representation for the gamma matrices.] The
fermion acquires different masses in the different protodomains due to its
coupling with $\varphi $, but the fermion becomes effectively massless in
the core of the domain wall where $\varphi \rightarrow 0$. It is therefore
energetically favorable for the fermion to reside inside the wall, and it is
assumed that fermions near the wall will become absorbed by the wall, so
that the wall is quickly populated by massless fermions. As an estimate, we
can think of a force of attraction acting on the fermions by the wall to be
given roughly by $f(x)\sim -\partial M(x)/\partial x\sim -G\partial \varphi
(x)/\partial x$. In the thin wall approximation, we can then regard the
domain wall as possessing a two dimensional Fermi gas of massless fermions.
The Fermi gas pressure can have a stabilizing influence by counteracting the
wall surface tension and false vacuum pressure contributions. For a
spherical vacuum bag, the collapse can be halted when the bag has a radius $%
R $, which minimizes the total energy $E$, and is related to the number of
fermions $N$ inhabiting the wall.

However, a vacuum bag of energy $E$ and radius $R$ is not stable against
flattening into a ``pancake'' shape. The tendency to flatten thus results in
the fragmentation of the vacuum bag into many smaller ones. The
fragmentation process halts when the thin wall approximation is no longer
valid, i.e. when the typical radius of curvature of a bag becomes comparable
to the wall thickness. In this limit, the configuration is better
represented by a ``Fermi ball'' which can be thought of as a tiny vacuum bag
with essentially no false vacuum in the interior - a nontopological scalar
field configuration consisting mostly of the domain wall inhabited by a
three dimensional Fermi gas. By equating the minimum size of the stabilized
Fermi ball $R_{\min }$ to the wall thickness $\delta $, and assuming the
stable Fermi ball to be spherical, the typical stabilizing radius at which
the collapse and fragmentation process stops is estimated to be 
\begin{equation}
R_{\min }\sim \frac 2{\lambda \varphi _0}.  \label{e40}
\end{equation}

The presence of a Fermi gas inside the domain wall is crucial to the
formation of Fermi balls in this model. In order that fermions and
antifermions do not undergo annihilation processes that leave no Fermi gas
inside the wall, it is sufficient to assume that there is a net fermion
antifermion asymmetry, so that annihilation processes which may occur will
eventually stop when all of the antifermions (or fermions) have been
consumed, leaving a Fermi gas of fermions (or antifermions).

MC estimate that a Fermi ball would contain about 50 fermions and have a
mass of roughly 100$\varphi _0$ GeV, where $\varphi _0$ is expressed in
units of GeV. If the Fermi balls are constructed from a new Dirac fermion
that has no standard model gauge charges, then the Fermi ball would likely
be a neutral, heavy, nonrelativistic particle interacting only very weakly
with ordinary matter. In this case, Fermi balls would form a candidate for
cold dark matter.

\subsection{Supersymmetry Breaking and Fermi Balls}

In the model presented here, the soft supersymmetry breaking terms are
embedded in the potential term $V_B$, given by (\ref{e35}). Thus, by (\ref
{e11}) and (\ref{e35}) the total Lagrangian, written in terms of the real
scalar fields $A$ and $B$ and the Majorana field $\Psi $, is given by 
\begin{equation}
\begin{array}{ll}
L_1 & =\partial ^\mu A\partial _\mu A+\partial ^\mu B\partial _\mu B+\frac i2%
\bar{\Psi}\gamma ^\mu \partial _\mu \Psi +i\lambda [A\bar{\Psi}\Psi +B\bar{%
\Psi}\gamma _5\Psi ] \\ 
& -\lambda ^2[(A^2-a^2)^2+2B^2(A^2+a^2)+B^4] \\ 
& -\mu ^2(A^2+B^2)-g_0(A^3-3AB^2)-V_0\,\,.
\end{array}
\label{e41}
\end{equation}

\noindent This model closely resembles that presented by MC, except that (%
\ref{e41}) contains an additional scalar field $B$ and the fermion here is
Majorana, rather than Dirac. [Let us also assume a parameter range that
keeps the vacuum expectation value $B_{vac}=0$ after supersymmetry breaking.
For example, we could require that $\frac{\mu ^2}{2\lambda ^2}<<a^2 $, $%
\frac{3g_0}{2\lambda ^2}<<a$, and $\mu ^2-3g_0a>0$, so that after the
breaking of the supersymmetry $B_{vac}=0$ and the vacuum values of $A$ are
only slightly shifted from $\pm a$.] Upon setting $B=0$, (\ref{e41}) is seen
to have the same form as (\ref{e39}). The supersymmetry breaking terms
embedded in $V_B$ also break the exact discrete symmetry associated with $%
A\rightarrow -A$ in the supersymmetric version. We therefore expect the
Fermi ball scenario to be realized in this broken supersymmetric model, as
well. Following the reasoning of MC, we expect a Fermi gas to remain within
the domain wall after possible fermion antifermion annihilations if there is
a fermion antifermion asymmetry, or when it becomes energetically
unfavorable for massive particles, like the $A$ and $B$ scalar bosons, to be
produced outside the wall. The Fermi balls associated with this model are
neutral, and again can be considered as candidates for cold dark matter. It
can also be pointed out that a slightly more complicated model\cite{mor},
composed of two interacting chiral superfields (and hence two Majorana
fermion fields, or equivalently, a Dirac field), that has the same basic
features presented here might be implemented. The Majorana fermion can then
be replaced with a Dirac fermion. It is conceivable that a model of this
type could be constructed where the scalar and spinor fields are allowed to
have standard model gauge couplings. But one of the points to be illuminated
in this work is that the MC type of model, which leads to the prediction of
Fermi balls, can emerge rather naturally in an initially supersymmetric
theory where the supersymmetry gets broken, along with the exact discrete
symmetry. The breaking of the exact discrete symmetry can remove the
potentially hazardous domain wall problem, and also give rise to the
production of Fermi balls.

\section{Summary}

Because there exists a strong possibility that (1) the early universe
underwent a set of symmetry breaking phase transitions, during which defects
may have been formed, and (2) supersymmetry was physically realized at the
time of defect production and was broken at a somewhat later time, it
becomes relevant to consider field theoretic models of defects within the
context of supersymmetry. Here, a simple domain wall model, constructed from
a single chiral superfield, has been examined. It is found that the fermions
become massless inside the domain wall, where a zero mode forms, and that
there are bound states describing real scalar bosons attached to the wall.
(These types of results have been examined previously for a supersymmetric
model constructed from two interacting chiral superfields\cite{mor}, but the
effects of soft supersymmetry breaking terms are more easily examined in the
single field model presented here.)

An exact $Z_2$ discrete symmetry in the bosonic sector of the theory,
associated with the reflection symmetry $A\rightarrow -A$, gets broken
explicitly to an approximate biased discrete symmetry when soft
supersymmetry breaking terms are added to the Lagrangian. The model then
closely resembles one describing ``Fermi balls'', which are scalar field
configurations stabilized by a Fermi gas exerting a degeneracy pressure. In
the context of the simple model presented here, we therefore expect the
production of domain walls, followed by a process wherein the walls form
false vacuum bags (due to the transformation of the exact discrete symmetry
into an approximate one), which collapse and fragment, finally resulting in
the production of Fermi balls. Thus, in the model presented here, the
production of Fermi balls is closely associated with the breaking of
supersymmetry.

\appendix 

\section{Conventions}

Some of the notations and conventions are briefly listed here. A metric $%
g_{\mu \nu }$ is used with signature $(+,-,-,-)$. Aside from the metric, the
notation, conventions, and gamma matrices used conform to those of ref.\cite
{sbook}. The gamma matrices can be written in the form 
\begin{equation}
\gamma ^\mu =i\left( 
\begin{array}{cc}
0 & \sigma ^\mu \\ 
\bar{\sigma}^\mu & 0
\end{array}
\right)  \label{a1}
\end{equation}

\noindent with 
\begin{equation}  \label{a2}
\sigma ^\mu =(1,{\bf \vec \sigma })\,\,,\,\,\,\,\,\,\,\,\,\,\bar \sigma ^\mu
=(1,-{\bf \vec \sigma })\,\,,
\end{equation}

\noindent where ${\bf \vec \sigma }$ represents the Pauli matrices. Then 
\begin{equation}  \label{a3}
\gamma ^0=i\left( 
\begin{array}{cc}
0 & 1 \\ 
1 & 0
\end{array}
\right) ,\,\,\,\,\,\,\,\,\,\,\gamma ^k=i\left( 
\begin{array}{cc}
0 & \sigma _k \\ 
-\sigma _k & 0
\end{array}
\right) ,\,\,\,\,\,k=1,2,3,
\end{equation}

\noindent and $\gamma _5$ is given by 
\begin{equation}  \label{a4}
\gamma _5=\gamma ^0\gamma ^1\gamma ^2\gamma ^3=i\left( 
\begin{array}{cc}
1 & 0 \\ 
0 & -1
\end{array}
\right) .
\end{equation}

\noindent The gamma matrices have the properties 
\begin{equation}  \label{a5}
\{\gamma ^\mu ,\gamma ^\nu \}=-2g^{\mu \nu },\,\,\,\,\{\gamma ^\mu ,\gamma
_5\}=0,\,\,\,\,\gamma _5^{\dagger }=-\gamma _5,\,\,\,\,(\gamma _5)^2=-1.
\end{equation}

\noindent A Majorana 4-spinor $\Psi $ is expressed in terms of the Weyl
2-spinors $\psi $ and $\bar \psi $ by $\Psi =\left( 
\begin{array}{c}
\psi _\alpha \\ 
\bar \psi ^{\dot \alpha }
\end{array}
\right) $ and we use the summation conventions for Weyl spinors [with $\bar 
\psi ^{\dot \alpha }=(\psi ^\alpha )^{*}$] 
\begin{equation}  \label{a6}
\xi \psi \equiv \xi ^\alpha \psi _\alpha ,\,\,\,\,\bar \xi \bar \psi \equiv 
\bar \xi _{\dot \alpha }\bar \psi ^{\dot \alpha },\,\,\,\,\alpha
=1,2,\,\,\,\,\dot \alpha =1,2,
\end{equation}

\noindent with $\varepsilon $ metric tensors (for raising and lowering Weyl
spinor indices) 
\begin{equation}
(\varepsilon ^{\alpha \beta })=(\varepsilon ^{\dot{\alpha}\dot{\beta}%
})=i\sigma _2,\,\,(\varepsilon _{\alpha \beta })=(\varepsilon _{\dot{\alpha}%
\dot{\beta}})=-i\sigma _2,\,\,\,\,\varepsilon ^{12}=1=\varepsilon ^{\dot{1}%
\dot{2}}.  \label{a7}
\end{equation}

\end{document}